\input harvmac

%\draftmode

%%%%%%%%%%%%%%%%% refs after 1806 %%%%%%%%%%%%%%

%\BabaroCMQ
\lref\BabaroCMQ{
  J.~P.~Babaro, V.~F.~Foit, G.~Giribet and M.~Leoni,
  ``$ T\overline{T} $ type deformation in the presence of a boundary,''
JHEP {\bf 1808}, 096 (2018).
[arXiv:1806.10713 [hep-th]].
%%CITATION = arXiv:1806.10713%%
}

%\ContiJHO
\lref\ContiJHO{
  R.~Conti, L.~Iannella, S.~Negro and R.~Tateo,
  ``Generalised Born-Infeld models, Lax operators and the $T\bar T$ perturbation,''
JHEP {\bf 1811}, 007 (2018).
[arXiv:1806.11515 [hep-th]].
%%CITATION = arXiv:1806.11515%%
}

%\ChenEQK
\lref\ChenEQK{
  B.~Chen, L.~Chen and P.~X.~Hao,
  ``Entanglement entropy in $T\overline{T}$-deformed CFT,''
Phys.\ Rev.\ D {\bf 98}, no. 8, 086025 (2018).
[arXiv:1807.08293 [hep-th]].
%%CITATION = arXiv:1807.08293%%
}

%\HartmanTKW
\lref\HartmanTKW{
  T.~Hartman, J.~Kruthoff, E.~Shaghoulian and A.~Tajdini,
  ``Holography at finite cutoff with a $T^2$ deformation,''
[arXiv:1807.11401 [hep-th]].
%%CITATION = arXiv:1807.11401%%
}

%\ChakrabortyAJI
\lref\ChakrabortyAJI{
  S.~Chakraborty,
  ``Wilson loop in a $T\bar{T}$ like deformed $\rm{CFT}_2$,''
[arXiv:1809.01915 [hep-th]].
%%CITATION = arXiv:1809.01915%%
}

%\CardyJHO
\lref\CardyJHO{
  J.~Cardy,
  ``$T\overline T$ deformations of non-Lorentz invariant field theories,''
[arXiv:1809.07849 [hep-th]].
%%CITATION = arXiv:1809.07849%%
}

%\ContiTCA
\lref\ContiTCA{
  R.~Conti, S.~Negro and R.~Tateo,
  ``The $T\bar T$ perturbation and its geometric interpretation,''
[arXiv:1809.09593 [hep-th]].
%%CITATION = arXiv:1809.09593%%
}

%\SantilliXUX
\lref\SantilliXUX{
  L.~Santilli and M.~Tierz,
  ``Large N phase transition in $ T\overline{T} $ -deformed 2d Yang-Mills theory on the sphere,''
[arXiv:1810.05404 [hep-th]].
%%CITATION = arXiv:1810.05404%%
}

%\BaggioRPV
\lref\BaggioRPV{
  M.~Baggio, A.~Sfondrini, G.~Tartaglino-Mazzucchelli and H.~Walsh,
  ``On $T\bar{T}$ deformations and supersymmetry,''
[arXiv:1811.00533 [hep-th]].
%%CITATION = arXiv:1811.00533%%
}

%\ChangDGE
\lref\ChangDGE{
  C.~K.~Chang, C.~Ferko and S.~Sethi,
  ``Supersymmetry and $T \overline{T}$ Deformations,''
[arXiv:1811.01895 [hep-th]].
%%CITATION = arXiv:1811.01895%%
}

%\NakayamaUJT
\lref\NakayamaUJT{
  Y.~Nakayama,
  ``Very Special $T\bar{J}$ deformed CFT,''
[arXiv:1811.02173 [hep-th]].
%%CITATION = arXiv:1811.02173%%
}

%\AraujoRHO
\lref\AraujoRHO{
  T.~Araujo, E.~Ó Colgáin, Y.~Sakatani, M.~M.~Sheikh-Jabbari and H.~Yavartanoo,
  ``Holographic integration of $T \bar{T}$ and $J \bar{T}$ via $O(d,d)$,''
[arXiv:1811.03050 [hep-th]].
%%CITATION = arXiv:1811.03050%%
}

%\WangJVA
\lref\WangJVA{
  P.~Wang, H.~Wu and H.~Yang,
  ``The dual geometries of $T\bar{T}$ deformed CFT$_2$ and highly excited states of CFT$_2$,''
[arXiv:1811.07758 [hep-th]].
%%CITATION = arXiv:1811.07758%%
}

%\GorbenkoOOV
\lref\GorbenkoOOV{
  V.~Gorbenko, E.~Silverstein and G.~Torroba,
  ``dS/dS and $T\bar T$,''
[arXiv:1811.07965 [hep-th]].
%%CITATION = arXiv:1811.07965%%
}

%\ParkSNF
\lref\ParkSNF{
  C.~Park,
  ``Holographic Entanglement Entropy in Cutoff AdS,''
[arXiv:1812.00545 [hep-th]].
%%CITATION = arXiv:1812.00545%%
}

%%%%%%%%%%%%%%%%% end of refs after 1806 %%%%%%%%%%

%%%%%%%%%%%%%%%%%% refs from jtbr paper %%%%%%%%%%%%%%%%%%%%

%\ElShowkCM
\lref\ElShowkCM{
  S.~El-Showk and M.~Guica,
  ``Kerr/CFT, dipole theories and nonrelativistic CFTs,''
JHEP {\bf 1212}, 009 (2012).
[arXiv:1108.6091 [hep-th]].
%%CITATION = arXiv:1108.6091%%
}

%\CompereJK
\lref\CompereJK{
  G.~Compère,
  ``The Kerr/CFT correspondence and its extensions,''
Living Rev.\ Rel.\  {\bf 15}, 11 (2012), [Living Rev.\ Rel.\  {\bf 20}, no. 1, 1 (2017)].
[arXiv:1203.3561 [hep-th]].
%%CITATION = arXiv:1203.3561%%
}

%\deBoerGYT
\lref\deBoerGYT{
  J.~de Boer, H.~Ooguri, H.~Robins and J.~Tannenhauser,
  ``String theory on AdS(3),''
JHEP {\bf 9812}, 026 (1998).
[hep-th/9812046].
%%CITATION = hep-th/9812046%%
}

%\GiveonKU
\lref\GiveonKU{
  A.~Giveon and A.~Pakman,
  ``More on superstrings in AdS(3) x N,''
JHEP {\bf 0303}, 056 (2003).
[hep-th/0302217].
%%CITATION = hep-th/0302217%%
}

%\BzowskiPCY
\lref\BzowskiPCY{
  A.~Bzowski and M.~Guica,
  ``The holographic interpretation of $J \bar T$-deformed CFTs,''
[arXiv:1803.09753 [hep-th]].
%%CITATION = arXiv:1803.09753%%
}

%\GiveonFU
\lref\GiveonFU{
  A.~Giveon, M.~Porrati and E.~Rabinovici,
  ``Target space duality in string theory,''
Phys.\ Rept.\  {\bf 244}, 77 (1994).
[hep-th/9401139].
%%CITATION = hep-th/9401139%%
}

%\ParsonsSI
\lref\ParsonsSI{
  J.~Parsons and S.~F.~Ross,
  ``Strings in extremal BTZ black holes,''
JHEP {\bf 0904}, 134 (2009).
[arXiv:0901.3044 [hep-th]].
%%CITATION = arXiv:0901.3044%%
}

%\BarsSR
\lref\BarsSR{
  I.~Bars and K.~Sfetsos,
  ``Conformally exact metric and dilaton in string theory on curved space-time,''
Phys.\ Rev.\ D {\bf 46}, 4510 (1992).
[hep-th/9206006].
%%CITATION = USC-92-HEP-B2%%
}

%\DetournayFZ
\lref\DetournayFZ{
  S.~Detournay, D.~Orlando, P.~M.~Petropoulos and P.~Spindel,
  ``Three-dimensional black holes from deformed anti-de Sitter,''
JHEP {\bf 0507}, 072 (2005).
[hep-th/0504231].
%%CITATION = hep-th/0504231%%
}

%\AzeyanagiZD
\lref\AzeyanagiZD{
  T.~Azeyanagi, D.~M.~Hofman, W.~Song and A.~Strominger,
  ``The Spectrum of Strings on Warped $AdS_3 \times S^3$,''
JHEP {\bf 1304}, 078 (2013).
[arXiv:1207.5050 [hep-th]].
%%CITATION = NSF-KITP-12-118%%
}

%\SmirnovLQW
\lref\SmirnovLQW{
  F.~A.~Smirnov and A.~B.~Zamolodchikov,
  ``On space of integrable quantum field theories,''
Nucl.\ Phys.\ B {\bf 915}, 363 (2017).
[arXiv:1608.05499 [hep-th]].
%%CITATION = arXiv:1608.05499%%
}

%\CavagliaODA
\lref\CavagliaODA{
  A.~Cavaglià, S.~Negro, I.~M.~Szécsényi and R.~Tateo,
  ``$T \bar{T}$-deformed 2D Quantum Field Theories,''
JHEP {\bf 1610}, 112 (2016).
[arXiv:1608.05534 [hep-th]].
%%CITATION = arXiv:1608.05534%%
}

%\McGoughLOL
\lref\McGoughLOL{
  L.~McGough, M.~Mezei and H.~Verlinde,
  ``Moving the CFT into the bulk with $ T\overline{T} $,''
JHEP {\bf 1804}, 010 (2018).
[arXiv:1611.03470 [hep-th]].
%%CITATION = arXiv:1611.03470%%
}

%\GiveonNIE
\lref\GiveonNIE{
  A.~Giveon, N.~Itzhaki and D.~Kutasov,
  ``$ T\bar{T} $ and LST,''
JHEP {\bf 1707}, 122 (2017).
[arXiv:1701.05576 [hep-th]].
%%CITATION = arXiv:1701.05576%%
}

%\PolchinskiRQ
\lref\PolchinskiRQ{
  J.~Polchinski,
  ``String theory. Vol. 1: An introduction to the bosonic string,''
}

%\GiveonMYJ
\lref\GiveonMYJ{
  A.~Giveon, N.~Itzhaki and D.~Kutasov,
  ``A solvable irrelevant deformation of AdS$_{3}$/CFT$_{2}$,''
JHEP {\bf 1712}, 155 (2017).
[arXiv:1707.05800 [hep-th]].
%%CITATION = arXiv:1707.05800%%
}

%\MaldacenaHW
\lref\MaldacenaHW{
  J.~M.~Maldacena and H.~Ooguri,
  ``Strings in AdS(3) and SL(2,R) WZW model 1: The Spectrum,''
J.\ Math.\ Phys.\  {\bf 42}, 2929 (2001).
[hep-th/0001053].
%%CITATION = hep-th/0001053%%
}

%\ShyamZNQ
\lref\ShyamZNQ{
  V.~Shyam,
  ``Background independent holographic dual to $T\bar{T}$ deformed CFT with large central charge in 2 dimensions,''
JHEP {\bf 1710}, 108 (2017).
[arXiv:1707.08118 [hep-th]].
%%CITATION = arXiv:1707.08118%%
}

%\AsratTZD
\lref\AsratTZD{
  M.~Asrat, A.~Giveon, N.~Itzhaki and D.~Kutasov,
  ``Holography Beyond AdS,''
[arXiv:1711.02690 [hep-th]].
%%CITATION = arXiv:1711.02690%%
}

%\GiribetIMM
\lref\GiribetIMM{
  G.~Giribet,
  ``$T\bar{T}$-deformations, AdS/CFT and correlation functions,''
JHEP {\bf 1802}, 114 (2018).
[arXiv:1711.02716 [hep-th]].
%%CITATION = arXiv:1711.02716%%
}

%\KrausXRN
\lref\KrausXRN{
  P.~Kraus, J.~Liu and D.~Marolf,
  ``Cutoff AdS$_3$ versus the $T\bar{T}$ deformation,''
[arXiv:1801.02714 [hep-th]].
%%CITATION = arXiv:1801.02714%%
}

%\CardySDV
\lref\CardySDV{
  J.~Cardy,
  ``The $T\overline T$ deformation of quantum field theory as a stochastic process,''
[arXiv:1801.06895 [hep-th]].
%%CITATION = arXiv:1801.06895%%
}

%\CottrellSKZ
\lref\CottrellSKZ{
  W.~Cottrell and A.~Hashimoto,
  ``Comments on $T \bar T$ double trace deformations and boundary conditions,''
[arXiv:1801.09708 [hep-th]].
%%CITATION = arXiv:1801.09708%%
}

%\AharonyVUX
\lref\AharonyVUX{
  O.~Aharony and T.~Vaknin,
  ``The TT* deformation at large central charge,''
[arXiv:1803.00100 [hep-th]].
%%CITATION = arXiv:1803.00100%%
}

%\DubovskyDLK
\lref\DubovskyDLK{
  S.~Dubovsky,
  ``A Simple Worldsheet Black Hole,''
[arXiv:1803.00577 [hep-th]].
%%CITATION = arXiv:1803.00577%%
}

%\BonelliKIK
\lref\BonelliKIK{
  G.~Bonelli, N.~Doroud and M.~Zhu,
  ``$T\bar T$-deformations in closed form,''
[arXiv:1804.10967 [hep-th]].
%%CITATION = arXiv:1804.10967%%
}

%\BaggioGCT
\lref\BaggioGCT{
  M.~Baggio and A.~Sfondrini,
  ``Strings on NS-NS Backgrounds as Integrable Deformations,''
[arXiv:1804.01998 [hep-th]].
%%CITATION = arXiv:1804.01998%%
}

%\ChakrabortyKPR
\lref\ChakrabortyKPR{
  S.~Chakraborty, A.~Giveon, N.~Itzhaki and D.~Kutasov,
  ``Entanglement Beyond $\rm AdS$,''
[arXiv:1805.06286 [hep-th]].
%%CITATION = arXiv:1805.06286%%
}

%\GuicaLIA
\lref\GuicaLIA{
  M.~Guica,
  ``An integrable Lorentz-breaking deformation of two-dimensional CFTs,''
[arXiv:1710.08415 [hep-th]].
%%CITATION = arXiv:1710.08415%%
}

%\BzowskiPCY
\lref\BzowskiPCY{
  A.~Bzowski and M.~Guica,
  ``The holographic interpretation of $J \bar T$-deformed CFTs,''
[arXiv:1803.09753 [hep-th]].
%%CITATION = arXiv:1803.09753%%
}

%\IRownNW
\lref\IRownNW{
  J.~D.~Brown and M.~Henneaux,
  ``Central Charges in the Canonical Realization of Asymptotic Symmetries: An Example from Three-Dimensional Gravity,''
Commun.\ Math.\ Phys.\  {\bf 104}, 207 (1986)..
}

%\GiveonCGS
\lref\GiveonCGS{
  A.~Giveon and D.~Kutasov,
  ``Supersymmetric Renyi entropy in CFT$_{2}$ and AdS$_{3}$,''
JHEP {\bf 1601}, 042 (2016).
[arXiv:1510.08872 [hep-th]].
%%CITATION = arXiv:1510.08872%%
}

%\KutasovXQ
\lref\KutasovXQ{
  D.~Kutasov and A.~Schwimmer,
  ``Universality in two-dimensional gauge theory,''
Nucl.\ Phys.\ B {\bf 442}, 447 (1995).
[hep-th/9501024].
%%CITATION = EFI-94-67%%
}

%\GiveonNS
\lref\GiveonNS{
  A.~Giveon, D.~Kutasov and N.~Seiberg,
  ``Comments on string theory on AdS(3),''
Adv.\ Theor.\ Math.\ Phys.\  {\bf 2}, 733 (1998).
[hep-th/9806194].
%%CITATION = hep-th/9806194%%
}

%\KutasovXU
\lref\KutasovXU{
  D.~Kutasov and N.~Seiberg,
  ``More comments on string theory on AdS(3),''
JHEP {\bf 9904}, 008 (1999).
[hep-th/9903219].
%%CITATION = hep-th/9903219%%
}

%\GiveonUP
\lref\GiveonUP{
  A.~Giveon and D.~Kutasov,
  ``Notes on AdS(3),''
Nucl.\ Phys.\ B {\bf 621}, 303 (2002).
[hep-th/0106004].
%%CITATION = RI-5-01%%
}

%\ArgurioTB
\lref\ArgurioTB{
  R.~Argurio, A.~Giveon and A.~Shomer,
  ``Superstrings on AdS(3) and symmetric products,''
JHEP {\bf 0012}, 003 (2000).
[hep-th/0009242].
%%CITATION = hep-th/0009242%%
}

%\GiveonMI
\lref\GiveonMI{
  A.~Giveon, D.~Kutasov, E.~Rabinovici and A.~Sever,
  ``Phases of quantum gravity in AdS(3) and linear dilaton backgrounds,''
Nucl.\ Phys.\ B {\bf 719}, 3 (2005).
[hep-th/0503121].
%%CITATION = hep-th/0503121%%
}

%\MotlTH
\lref\MotlTH{
  L.~Motl,
  ``Proposals on nonperturbative superstring interactions,''
[hep-th/9701025].
%%CITATION = hep-th/9701025%%
}

%\DijkgraafVV
\lref\DijkgraafVV{
  R.~Dijkgraaf, E.~P.~Verlinde and H.~L.~Verlinde,
  ``Matrix string theory,''
Nucl.\ Phys.\ B {\bf 500}, 43 (1997).
[hep-th/9703030].
%%CITATION = hep-th/9703030%%
}

%\GiveonCGS
\lref\GiveonCGS{
  A.~Giveon and D.~Kutasov,
  ``Supersymmetric Renyi entropy in CFT$_{2}$ and AdS$_{3}$,''
JHEP {\bf 1601}, 042 (2016).
[arXiv:1510.08872 [hep-th]].
%%CITATION = arXiv:1510.08872%%
}

%\KutasovXB
\lref\KutasovXB{
  D.~Kutasov,
  ``Geometry on the Space of Conformal Field Theories and Contact Terms,''
Phys.\ Lett.\ B {\bf 220}, 153 (1989).
%%CITATION = WIS-88-55-PH%%
}

%\GiveonZM
\lref\GiveonZM{
  A.~Giveon, D.~Kutasov and O.~Pelc,
  ``Holography for noncritical superstrings,''
JHEP {\bf 9910}, 035 (1999).
[hep-th/9907178].
%%CITATION = hep-th/9907178%%
}

%\ZamolodchikovBD
\lref\ZamolodchikovBD{
  A.~B.~Zamolodchikov and V.~A.~Fateev,
  ``Operator Algebra and Correlation Functions in the Two-Dimensional Wess-Zumino SU(2) x SU(2) Chiral Model,''
Sov.\ J.\ Nucl.\ Phys.\  {\bf 43}, 657 (1986), [Yad.\ Fiz.\  {\bf 43}, 1031 (1986)].
}

%\TeschnerFT
\lref\TeschnerFT{
  J.~Teschner,
  ``On structure constants and fusion rules in the SL(2,C) / SU(2) WZNW model,''
Nucl.\ Phys.\ B {\bf 546}, 390 (1999). [hep-th/9712256]; A. B. Zamolodchikov and Al. B.
Zamolodchikov, unpublished.
%%CITATION = hep-th/9712256%%
}

%\KlemmDF
\lref\KlemmDF{
  A.~Klemm and M.~G.~Schmidt,
  ``Orbifolds by Cyclic Permutations of Tensor Product Conformal Field Theories,''
Phys.\ Lett.\ B {\bf 245}, 53 (1990).
%%CITATION = HD-THEP-90-13%%
}

%\FuchsVU
\lref\FuchsVU{
  J.~Fuchs, A.~Klemm and M.~G.~Schmidt,
  ``Orbifolds by cyclic permutations in Gepner type superstrings and in the corresponding Calabi-Yau manifolds,''
Annals Phys.\  {\bf 214}, 221 (1992).
%%CITATION = HD-THEP-90-34%%
}

%\WakimotoGF
\lref\WakimotoGF{
  M.~Wakimoto,
  ``Fock representations of the affine lie algebra A1(1),''
Commun.\ Math.\ Phys.\  {\bf 104}, 605 (1986).
}

%\BernardIY
\lref\BernardIY{
  D.~Bernard and G.~Felder,
  ``Fock Representations and BRST Cohomology in SL(2) Current Algebra,''
Commun.\ Math.\ Phys.\  {\bf 127}, 145 (1990).
%%CITATION = SACLAY-SPH-T-89-113%%
}

%\BershadskyIN
\lref\BershadskyIN{
  M.~Bershadsky and D.~Kutasov,
  ``Comment on gauged WZW theory,''
Phys.\ Lett.\ B {\bf 266}, 345 (1991).
%%CITATION = PUPT-1261%%
}

%\ElitzurRT
\lref\ElitzurRT{
  S.~Elitzur, A.~Giveon, D.~Kutasov and E.~Rabinovici,
  ``From big bang to big crunch and beyond,''
JHEP {\bf 0206}, 017 (2002).
[hep-th/0204189].
%%CITATION = hep-th/0204189%%
}

%\CrapsII
\lref\CrapsII{
  B.~Craps, D.~Kutasov and G.~Rajesh,
  ``String propagation in the presence of cosmological singularities,''
JHEP {\bf 0206}, 053 (2002).
[hep-th/0205101].
%%CITATION = hep-th/0205101%%
}

%\ZamolodchikovCE
\lref\ZamolodchikovCE{
  A.~B.~Zamolodchikov,
  ``Expectation value of composite field T anti-T in two-dimensional quantum field theory,''
[hep-th/0401146].
%%CITATION = hep-th/0401146%%
}

%\CaselleDRA
\lref\CaselleDRA{
  M.~Caselle, D.~Fioravanti, F.~Gliozzi and R.~Tateo,
  ``Quantisation of the effective string with TBA,''
JHEP {\bf 1307}, 071 (2013).
[arXiv:1305.1278 [hep-th]].
%%CITATION = arXiv:1305.1278%%
}

%\DattaTHY
\lref\DattaTHY{
  S.~Datta and Y.~Jiang,
  ``$T\bar{T}$ deformed partition functions,''
[arXiv:1806.07426 [hep-th]].
%%CITATION = arXiv:1806.07426%%
}

%\IsraelVV
\lref\IsraelVV{
  D.~Israel, C.~Kounnas, D.~Orlando and P.~M.~Petropoulos,
  ``Electric/magnetic deformations of S**3 and AdS(3), and geometric cosets,''
Fortsch.\ Phys.\  {\bf 53}, 73 (2005).
[hep-th/0405213].
%%CITATION = hep-th/0405213%%
}

%\DubovskyIRA
\lref\DubovskyIRA{
  S.~Dubovsky, V.~Gorbenko and M.~Mirbabayi,
  ``Natural Tuning: Towards A Proof of Concept,''
JHEP {\bf 1309}, 045 (2013).
[arXiv:1305.6939 [hep-th]].
%%CITATION = arXiv:1305.6939%%
}

%\DubovskyCNJ
\lref\DubovskyCNJ{
  S.~Dubovsky, V.~Gorbenko and M.~Mirbabayi,
  ``Asymptotic fragility, near AdS$_{2}$ holography and $ T\overline{T} $,''
JHEP {\bf 1709}, 136 (2017).
[arXiv:1706.06604 [hep-th]].
%%CITATION = arXiv:1706.06604%%
}

%\ApoloQPQ
\lref\ApoloQPQ{
  L.~Apolo and W.~Song,
  ``Strings on warped AdS$_3$ via $T\bar{J}$ deformations,''
[arXiv:1806.10127 [hep-th]].
%%CITATION = arXiv:1806.10127%%
}

%\DetournayRH
\lref\DetournayRH{
  S.~Detournay, D.~Israel, J.~M.~Lapan and M.~Romo,
  ``String Theory on Warped $AdS_{3}$ and Virasoro Resonances,''
JHEP {\bf 1101}, 030 (2011).
[arXiv:1007.2781 [hep-th]].
%%CITATION = arXiv:1007.2781%%
}

%\AharonyBAD
\lref\AharonyBAD{
  O.~Aharony, S.~Datta, A.~Giveon, Y.~Jiang and D.~Kutasov,
  ``Modular invariance and uniqueness of $T\bar{T}$ deformed CFT,''
[arXiv:1808.02492 [hep-th]].
%%CITATION = arXiv:1808.02492%%
}

%\AharonyICS
\lref\AharonyICS{
  O.~Aharony, S.~Datta, A.~Giveon, Y.~Jiang and D.~Kutasov,
  ``Modular covariance and uniqueness of $J\bar{T}$ deformed CFTs,''
[arXiv:1808.08978 [hep-th]].
%%CITATION = arXiv:1808.08978%%
}

%%%%%%%%%%%%%%%%%%%%% end of refs fron jtbar paper %%%%%%%%%%%%%%%%%%%%%%%

%\GuicaLIA
\lref\GuicaLIA{
  M.~Guica,
  ``An integrable Lorentz-breaking deformation of two-dimensional CFTs,''
SciPost Phys.\  {\bf 5}, no. 5, 048 (2018).
[arXiv:1710.08415 [hep-th]].
%%CITATION = arXiv:1710.08415%%
}

%\ApoloQPQ
\lref\ApoloQPQ{
  L.~Apolo and W.~Song,
  ``Strings on warped AdS$_{3}$ via $ T\bar J$ deformations,''
JHEP {\bf 1810}, 165 (2018).
[arXiv:1806.10127 [hep-th]].
%%CITATION = arXiv:1806.10127%%
}

%\ParsonsSI
\lref\ParsonsSI{
  J.~Parsons and S.~F.~Ross,
  ``Strings in extremal BTZ black holes,''
JHEP {\bf 0904}, 134 (2009).
[arXiv:0901.3044 [hep-th]].
%%CITATION = arXiv:0901.3044%%
}

%\KutasovXU
\lref\KutasovXU{
  D.~Kutasov and N.~Seiberg,
  ``More comments on string theory on AdS(3),''
JHEP {\bf 9904}, 008 (1999).
[hep-th/9903219].
%%CITATION = hep-th/9903219%%
}

%\AzeyanagiZD
\lref\AzeyanagiZD{
  T.~Azeyanagi, D.~M.~Hofman, W.~Song and A.~Strominger,
  ``The Spectrum of Strings on Warped $AdS_3\times S^3$,''
JHEP {\bf 1304}, 078 (2013).
[arXiv:1207.5050 [hep-th]].
%%CITATION = NSF-KITP-12-118%%
}

%\GiveonFU
\lref\GiveonFU{
  A.~Giveon, M.~Porrati and E.~Rabinovici,
  ``Target space duality in string theory,''
Phys.\ Rept.\  {\bf 244}, 77 (1994).
[hep-th/9401139].
%%CITATION = hep-th/9401139%%
}

%\ForsteWP
\lref\ForsteWP{
  S.~Forste,
  ``A Truly marginal deformation of SL(2, R) in a null direction,''
Phys.\ Lett.\ B {\bf 338}, 36 (1994).
[hep-th/9407198].
%%CITATION = hep-th/9407198%%
}

%\MaldacenaHW
\lref\MaldacenaHW{
  J.~M.~Maldacena and H.~Ooguri,
  ``Strings in AdS(3) and SL(2,R) WZW model 1.: The Spectrum,''
J.\ Math.\ Phys.\  {\bf 42}, 2929 (2001).
[hep-th/0001053].
%%CITATION = hep-th/0001053%%
}

%\ArgurioTB
\lref\ArgurioTB{
  R.~Argurio, A.~Giveon and A.~Shomer,
  ``Superstrings on AdS(3) and symmetric products,''
JHEP {\bf 0012}, 003 (2000).
[hep-th/0009242].
%%CITATION = hep-th/0009242%%
}

%\GiveonMI
\lref\GiveonMI{
  A.~Giveon, D.~Kutasov, E.~Rabinovici and A.~Sever,
  ``Phases of quantum gravity in AdS(3) and linear dilaton backgrounds,''
Nucl.\ Phys.\ B {\bf 719}, 3 (2005).
[hep-th/0503121].
%%CITATION = hep-th/0503121%%
}

%\GiveonNIE
\lref\GiveonNIE{
  A.~Giveon, N.~Itzhaki and D.~Kutasov,
  ``$T\bar T$ and LST,''
JHEP {\bf 1707}, 122 (2017).
[arXiv:1701.05576 [hep-th]].
%%CITATION = arXiv:1701.05576%%
}

%\GiveonMYJ
\lref\GiveonMYJ{
  A.~Giveon, N.~Itzhaki and D.~Kutasov,
  ``A solvable irrelevant deformation of AdS$_{3}$/CFT$_{2}$,''
JHEP {\bf 1712}, 155 (2017).
[arXiv:1707.05800 [hep-th]].
%%CITATION = arXiv:1707.05800%%
}
%

%\ChakrabortyVJA
\lref\ChakrabortyVJA{
  S.~Chakraborty, A.~Giveon and D.~Kutasov,
  ``$ J\overline{T} $ deformed CFT$_{2}$ and string theory,''
JHEP {\bf 1810}, 057 (2018).
[arXiv:1806.09667 [hep-th]].
%%CITATION = arXiv:1806.09667%%
}

\lref\cdgjk{
S.~Chakraborty, S.~Datta, A.~Giveon, Y.~Jiang and D.~Kutasov, work in progress.
}

%\AharonyICS
\lref\AharonyICS{
  O.~Aharony, S.~Datta, A.~Giveon, Y.~Jiang and D.~Kutasov,
  ``Modular covariance and uniqueness of $J\bar{T}$ deformed CFTs,''
[arXiv:1808.08978 [hep-th]].
%%CITATION = arXiv:1808.08978%%
}

%\AharonyBAD
\lref\AharonyBAD{
  O.~Aharony, S.~Datta, A.~Giveon, Y.~Jiang and D.~Kutasov,
  ``Modular invariance and uniqueness of $T\bar{T}$ deformed CFT,''
[arXiv:1808.02492 [hep-th]].
%%CITATION = arXiv:1808.02492%%
}

%\SmirnovLQW
\lref\SmirnovLQW{
  F.~A.~Smirnov and A.~B.~Zamolodchikov,
  ``On space of integrable quantum field theories,''
Nucl.\ Phys.\ B {\bf 915}, 363 (2017).
[arXiv:1608.05499 [hep-th]].
%%CITATION = arXiv:1608.05499%%
}

%\CavagliaODA
\lref\CavagliaODA{
  A.~Cavaglià, S.~Negro, I.~M.~Szécsényi and R.~Tateo,
  ``$T \bar{T}$-deformed 2D Quantum Field Theories,''
JHEP {\bf 1610}, 112 (2016).
[arXiv:1608.05534 [hep-th]].
%%CITATION = arXiv:1608.05534%%
}

%\SonYE
\lref\SonYE{
  D.~T.~Son,
  ``Toward an AdS/cold atoms correspondence: A Geometric realization of the Schrodinger symmetry,''
Phys.\ Rev.\ D {\bf 78}, 046003 (2008).
[arXiv:0804.3972 [hep-th]].
%%CITATION = arXiv:0804.3972%%
}

%\BalasubramanianDM
\lref\BalasubramanianDM{
  K.~Balasubramanian and J.~McGreevy,
  ``Gravity duals for non-relativistic CFTs,''
Phys.\ Rev.\ Lett.\  {\bf 101}, 061601 (2008).
[arXiv:0804.4053 [hep-th]].
%%CITATION = arXiv:0804.4053%%
}

\lref\blfmm{B. Le Floch and M. Mezei,
``Solving a family of $T\bar T$-like theories,'' to appear in a coordinated submission to the arXiv.
}

%%%%%%%%%%%%%%%%%%%%%%%%%%%%%%%%%%%%%%%%%%%%%%%%%%%
\Title{} {\centerline{Comments on $T\bar T$, $J\bar{T}$ and String Theory}}

\bigskip
\centerline{\it Amit Giveon}
\bigskip
\smallskip
\centerline{Racah Institute of Physics, The Hebrew
University} \centerline{Jerusalem 91904, Israel}

\smallskip

\vglue .3cm

\bigskip

\bigskip
\noindent
These notes contain some aspects of the holographic duality between
the perturbative superstring on current-current deformations of $AdS_3\times S^1\times{\cal N}$
and single-trace $T\bar T$ and $J\bar T$ deformed $CFT_2$.~\foot{Lecture given at
the 36th Jerusalem Winter School in Theoretical Physics,
Recent Progress in Quantum Field/String Theory,
December 30 -- January 10, 2019.}

\bigskip

\Date{12/18}

%%%%%%%%%%%%%%%%%%%%%%%%%%%%%%%%%%%%%%%%%%%%%%%%%%%%%%%%%%%%%%%%

\newsec{Introduction}

Perturbative string theory on {\it solvable} worldsheet CFT backgrounds
-- {\it current-current deformations of the superstring on $AdS_3\times S^1\times\cal N$} --
allows one to conjecture a holographic duality with {\it solvable irrelevant deformations} of $CFT_2$,
e.g. `single-trace' $T\bar T$ and $J\bar T$ deformed $CFT_2$;
we shall describe some aspects of these in the next sections.

Motivation:
\item{$\bullet$}
This construction provides holographic duals for a large class of vacua of string theory in asymptotically flat
linear dilaton spacetimes,
and sheds light on the UV behavior of $T\bar T$ deformed $CFT_2$;~\foot{In terms of {\it non-local} theories.}
it may provide a {\it step towards holography in flat spacetime}.
\item{$\bullet$}
The holography above is likely to have interesting consequences for the physics
of {\it horizons/singularities/closed timelike curves (CTC's) in string theory}.

Comments:
\item{*}
By a `single-trace' (ST) deformation, in the context of this note,
we mean e.g. a deformation in the block ${\cal M}$ of $Symm^N({\cal M})$;
its precise definition is the deformation of the boundary $CFT_2$ dual
to the perturbative superstring theory on the deformed $AdS_3\times S^1\times{\cal N}$.~\foot{By `$AdS_3$' we mean either $SL(2,R)$,
and/or global $AdS_3$ and/or $\{$massless BTZ$\}$=$\{AdS_3$ in Poincar\'e coordinates, with a compact boundary spatial direction$\}$,
and/or $H_3^+$.}
\item{*}
The holographic conjecture `predicted,' in particular,~\foot{More predictions are e.g.
those concerning the entanglement entropy in various cases.}  the spectrum of $J\bar T$ deformed $CFT_2$,
which was confirmed in field theory (with reasonable assumptions).
\item{*}
We also have~\cdgjk\ a holographic duality conjecture {\it predicting the spectrum of a $CFT_2$ deformed by $-tT\bar T+\mu J\bar T$}
(which is partly confirmed with reasonable assumptions).
Concretely,~\foot{One may skip the concrete prediction; it will be discuseed in section 9.}
the energy $E(t,\mu;R)$ and momentum $P(R)$
of states in the deformed $CFT_2$, on a cylinder with radius $R$, are given in terms of
the left and right-handed scaling dimensions of the states in the
the original theory, $h,\bar h$, the original holomorphic $U(1)$ charge, $q$,
and the central charge $c$ of the $CFT_2$, by
\eqn\epred{ER=n+{1\over 2A}\left(-B-\sqrt{B^2-4AC}\right);\qquad PR=n~,}
where
\eqn\abcpred{A={1\over 16\pi R^2}\left(\pi\mu^2-8t\right),\quad B=-1+{\mu\over 2R}q-{t\over\pi R^2}n,
\quad C=2\left(\bar h-{c\over 24}\right); \quad n=h-\bar h~.}

The plan of this note is the following.
In section 2, we recall some `experimental' results and comment on the qualitative reasons
that the spectrum of the perturbative superstring on $AdS_3$ has the pattern of a symmetric orbifold $CFT_2$,
and in section 3, we recall how it is obtained in the Ramond sector of the boundary $CFT_2$, namely, for the superstring
on the massless BTZ worldsheet background.
In section 4,~\foot{It is possible to read section 4 prior to sections 2,3
(and then to read section 3 before 2).}
we present a heuristic argument for a holographic duality
between the superstring on certain backgrounds,
e.g. on
\eqn\mthree{{\cal M}_3\equiv J^-\bar J^-\, {\rm deformed}\, AdS_3~,}
and ST-$T\bar T$ deformed $CFT_2$,~\foot{`ST-$T\bar T$' denotes `Single-Trace $T\bar T$,'
namely, $T\bar T$ in perturbative String Theory (and, similarly, for other deformations below).}
and in sections 5 and 6, we recall some properties of the ${\cal M}_3$ $CFT_2$ background
and the spectrum of the superstring on \mthree, respectively.
In sections 7 and 8, we conjecture, similarly,
a holographic duality between the superstring on $WAdS_3$ and ST-$J\bar T$ deformed $CFT_2$,
where $WAdS_3$ is the KK reduction to $3d$ of
\eqn\wadsthree{K\bar J^-\, {\rm deformed}\, AdS_3\times S^1~,}
with $K(z)$ being the $U(1)$ holomorphic current of the $S^1$ $CFT_2$.
Finally, in section 9,~\foot{One may read section 9 before sections 7,8;
the latter, as well as sections 5,6,
can be regarded as comments on special cases of section 9.}
we present some preliminary results concerning a holographic duality conjecture between
the superstring on a familiy of backgrounds, combining \mthree\ and \wadsthree, namely,
\eqn\wadsthree{\lambda J^-\bar J^-+\epsilon K\bar J^-\, {\rm deformed}\, AdS_3\times S^1~,}
and ST-$\{-tT\bar T+\mu J\bar T\}$ deformed $CFT_2$,
and in section 10, we list some open problems.

There is a handful of works on $T\bar T$, $J\bar T$ and string theory; see e.g.
\refs{\ZamolodchikovCE\CaselleDRA\DubovskyIRA\SmirnovLQW\CavagliaODA\McGoughLOL\GiveonNIE\DubovskyCNJ\GiveonMYJ\ShyamZNQ\GuicaLIA\AsratTZD\GiribetIMM\KrausXRN\CardySDV\CottrellSKZ\AharonyVUX\DubovskyDLK\BzowskiPCY\BonelliKIK\BaggioGCT\ChakrabortyKPR\DattaTHY\ChakrabortyVJA\ApoloQPQ\BabaroCMQ\ContiJHO\ChenEQK\HartmanTKW\AharonyBAD\AharonyICS\ChakrabortyAJI\CardyJHO\ContiTCA\SantilliXUX\BaggioRPV\ChangDGE\NakayamaUJT\AraujoRHO\WangJVA\GorbenkoOOV-\ParkSNF}.

The main results in sections 2--8 are based on \refs{\GiveonNIE,\GiveonMYJ,\ChakrabortyVJA,\AharonyBAD,\AharonyICS}
and references therein;
in particular, many comments in this file are taken from there.
The study in sections 7 and 8 was motivated by \GuicaLIA\ and was done independently in \ApoloQPQ.

On the other hand, the results in section 9 are new \cdgjk.

{\bf Note added}: The new results in \epred,\abcpred\ were obtained independently in \blfmm.
% We thank the authors for comparing our formulas.

\newsec{Comments on superstring theory on $AdS_3$ and ${\cal M}^N/S_N$}

%from section 5 of zamo-ws-st

The spectrum of the boundary $CFT_2$ on the cylinder
dual to the perturbtive superstring theory~\foot{By `perturbative string theory,' we mean a theory with a worldsheet $CFT_2$
background (in particular, in the case of a sigma-model, it involves only NS-NS background fields),
and with a parmetrically small string coupling, $g_s\ll 1$; below, by `string theory,'
we mean a perturbative string theory of this type.}
on $AdS_3$ has the structure of ${\cal M}^N/S_N$;~\foot{At large $N$,
namely, up to corrections in $g_s^2\sim 1/N$.}
%both in the Ramond (R) sector and the Neveu-Schwarz (NS) sector;
here we list some known results
regarding this pattern of the spectrum, and comment on them:
\item{*}
The Ramond sector of the dual $CFT_2$ on the cylinder
is given by the superstring on $M=0$ BTZ; it will be described in the next section.
\item{*}
The spectrum of the superstring on $M=0$ BTZ was obtained e.g. in section 5.1 of \ChakrabortyVJA,
and is the same as that of a symmetric product CFT, ${\cal M}^N/S_N$, where the block ${\cal M}$
is the $c=6k$ CFT associated with one string in the BTZ background and the winding $w$ labels the twisted sector;
it will be described in the next section.
\item{*}
The qualitative reason to expect a relation between the string spectrum and the symmetric orbifold is the following.
The states in the type II superstring on massless BTZ$\times\cal N$
can be thought of as describing strings moving in the radial direction in a particular state of excitation.
These strings are free (at large $N$,~\foot{$N$ is the number of BPS fundamental strings with their worldvolume in the
$R_t\times S^1$ of a linear dilaton background,
$R_t\times S^1\times R_\phi\times\cal N$,
whose near-horizon geometry is $M=0$ BTZ$\times\cal N$.}
or small string coupling, $g_s^2\sim 1/N$),
and the fact that they can be described by a symmetric product is very similar to that utilized in matrix string theory.
\item{*}
In different words, the background corresponding to the boundary $CFT_2$ on the cylinder
in its Ramond sector (i.e. with unbroken supersymmetry on the cylinder),
is obtained by replacing $AdS_3$ by the $M=J=0$ BTZ black hole.
The strings and fivebranes~\foot{Generically, with the NS5-branes wrapped around a singular
four-cycle in a Calabi-Yau four-fold and/or considering generic supersymetric linear dilaton throats,
instead of fivebranes, as in section 4.}
that create the background are mutually BPS in this case.
Thus, their potential is flat.
This means that there is a continuum of states corresponding to strings moving radially away from the fivebranes.
These states are described by a symmetric product, as in matrix string theory.
\item{*}
The NS sector of the dual $CFT_2$ on the cylinder is given by the superstring on global $AdS_3$.
The perturbative superstring spectrum in this sector
was investigated in the superstring followups of \MaldacenaHW, e.g., \refs{\ArgurioTB,\GiveonMI};
for states in the continuous representations,~\foot{And discrete states that amount to chiral states in the boundary $CFT_2$.}
it has the same structure as that of a symmetric product CFT, ${\cal M}^N/S_N$, where the block ${\cal M}$
is the $c=6k$ CFT associated with one string in the $AdS_3$ background and the winding $w$ labels the twisted sector.
\item{*}
The heuristic arguments concerning the symmetric product structure in the R sector of the theory
do not apply directly to the NS sector, e.g. since the discrete states live in the interior of $AdS_3$.~\foot{The study
of spacetime properties of such states from perturbative string theory is subtle,
since, generically, only a very small fraction of them
arise directly from principal discrete representations of $SL(2,R)$;
their vast majority is `hidden' in the continuum. Nevertheless, it will be interesting to investigate the properties
of on-shell discrete string states that can be seen in perturbative string theoy, beyond the chiral ones.}
%nevertheless, the spectrum has the structure of ${\cal M}^N/S_N$.~\foot{Perhaps the reason that this
%`must have been the case' is a consequence of modular invariance of the spacetime $CFT_2$.}

\newsec{Superstring theory on $M=0$ BTZ}

In this section, we study the spectrum of perturbative string theory on the massless BTZ background.
We start, in subsection 3.1, by reviewing the spectrum of the worldsheet theory,
and in subsection 3.2, we review the spectrum of the spacetime theory.

\subsec{The spectrum of the worldsheet theory}

%following section 5.1.1 in CGK1806

The worldsheet sigma-model Lagrangian on $M=0$ BTZ is~\foot{I am very telegraphic here; more details can be found
in \ChakrabortyVJA\ and references therein.}
\eqn\wssbtz{{\cal L}=2k\left(\partial\phi\bar\partial\phi+e^{2\phi}\partial\bar\gamma\bar\partial\gamma\right)~,}
where
\eqn\gggggr{\gamma=\gamma_1+\gamma_0~,\quad\bar\gamma=\gamma_1-\gamma_0~;\qquad\gamma_1\simeq\gamma_1+2\pi R~.}
It is convenient to rewrite it in the Wakimoto form~\foot{Treating carefully the measure of the path integral,
while integrating out $\beta,\bar\beta$,
and rescaling the fields, gives back the Lagrangian in \wssbtz.}
\eqn\wakimoto{{\cal L}_W=\beta\bar\partial\gamma+\bar\beta\partial\bar\gamma+
\partial\phi\bar\partial\phi-\sqrt{2\over k}{\hat R}\phi-e^{-\sqrt{2\over k}\phi}\beta\bar\beta~.}
Near the boundary, ${\cal L}_W$ becomes free,
\eqn\lfree{{\cal L}=\beta\bar\partial\gamma+\bar\beta\partial\bar\gamma+{\cal L}_\phi~,}
where
\eqn\lphi{{\cal L}_\phi=\partial\phi\bar\partial\phi-\sqrt{2\over k}{\hat R}\phi~,}
and it is thus useful for calculations.

It will be further useful for us, following \ParsonsSI, to bosonize the $\beta$-$\gamma$ system as
\eqn\phipm{\gamma=i\phi_-,\,\, \bar\gamma=i\bar\phi_-~;\qquad\beta=i\partial\phi_+,\,\, \bar\beta=i\bar\partial\bar\phi_+~,}
where $\phi_\pm$ are canonically normalized light-like fields,
\eqn\phiphi{\phi_+(z)\phi_-(w)\sim\ln(z-w)~.}
In terms of $\phi_\pm$, \lfree\ takes the form
\eqn\lphipm{{\cal L}=-\partial\phi_+\bar\partial\phi_--\partial\phi_-\bar\partial\phi_++{\cal L}_\phi~.}
To obtain the massless BTZ orbifolding, one introduces the twist fields,
\eqn\twist{t^w=e^{iw(\phi_++\bar\phi_+)}~,\qquad w\in Z~,}
and imposes mutual locality w.r.t. to $t^w$ as well as adding the $w$ twisted sectors
via OPE's of the untwisted operators with \twist.~\foot{W.l.g. we consider states with $w>0$;
those with negative $w$ amount to their conjugates (namely, outgoing versus ingoing or annihilation versus creation).}
The vertex operators thus obtained take the form~\foot{We present only those of interest, for simplicity.}
\eqn\vbtz{V_{BTZ}=e^{\sqrt{2\over k}j(\phi+\bar\phi)}V^w_{E_{L,R}}~,}
where
\eqn\vwelr{V^w_{E_{L,R}}=e^{iw\phi_++iE_L\phi_-}e^{iw\bar\phi_++iE_R\bar\phi_-}~,}
with
\eqn\elerp{(E_L,E_R)={R\over 2}(E+P,E-P)~;\qquad P={n\over R}~,\,\, n\in Z~.}

Comments:
\item{*}
$R$ sets the scale of the problem;
one can think of it as the radius of compactification of the geometric coordinate $\gamma_1$, \gggggr.
\item{*}
In the Wakimoto representation, one has
\eqn\jminus{J^-=\beta=i\partial\phi_+~;\quad \bar J^-=\bar\beta=i\bar\partial\phi_+~,}
where $J^-$ ($\bar J^-$) is the null holomorphic (anti-holomorphic) current of the affine $SL(2,R)_L$ ($SL(2,R)_R$)
symmetry of the underlying $SL(2,R)$ WZW model
(\wssbtz\ prior to compactifying $\gamma_1$, namely, $AdS_3$ in Poincar\'e coordinates).
\item{*}
$V_{BTZ}$, \vbtz, thus correspond to eigenstates of $J^-$ and $\bar J^-$ with eigenvalues $E_L$ and $E_R$,
respectively.
\item{*}
$E$ and $P$ are the energy and momentum of these states, respectively.
\item{*}
States which carry real radial momentum $P_\phi$ have
\eqn\jjss{j=-{1\over 2}+is~,\quad s\in R~,}
with $s$ proportional to $P_\phi$; they amount to the principal continuous representations of the underlying
$SL(2,R)$.~\foot{In the perturbative superstring on global $AdS_3$, \refs{\MaldacenaHW,\ArgurioTB},
one can construct physical, discrete bound states in the principal discrete representations of $SL(2,R)$;
in massless BTZ, the eigenvalues of $J^-$ can take any real value for both the continuous and discrete representations,
and thus all the states are part of a continuum.}
\item{*}
The integer $w$ labels different twisted sectors.~\foot{One can think of $w$ as the winding number of the string
around the circle on the boundary of BTZ; this is particularly clear in the semiclassical approach of section 4 in \ChakrabortyVJA.}
These sectors are constructed in a way analogous to \refs{\MaldacenaHW,\ArgurioTB},
but here the spectral flow/twist is in the $J^-$ direction,
whereas there it was in the $J^3$ direction.

The left and right-handed scaling dimensions $\Delta_{L,R}$ of $V_{BTZ}$ are
\eqn\deltadelta{\Delta_{L,R}=-wE_{L,R}-{j(j+1)\over k}~,\qquad \Delta_R-\Delta_L=wn~.}
We will next use them to calculate the spectrum of the spacetime theory.

\subsec{The spectrum of the spacetime theory}

%following section 5.1.2 in CGK1806

Consider the type II superstring on massless BTZ$\times{\cal N}$, which corresponds to a Ramond ground state of the
dual $CFT_2$. Let
\eqn\letvphys{V_{phys}=e^{-\varphi-\bar\varphi}V_{BTZ}V_{\cal N}}
be a physical vertex operator of the theory.~\foot{This is
only a particular class of operators, which is sufficient for simplicity.}
The on-shell condition reads:
\eqn\onshellbtz{\Delta_{L,R}+N_{L,R}-{1\over 2}=0~,}
where $N_{L,R}$ are the left and right-handed scaling dimensions of $V_{\cal N}$.~\foot{Note that \onshellbtz\ with \deltadelta,
and assuming that the GSO projection eliminates the tachyon,
imply, in particular, that states in the $w=0$ sector are necessarily discrete states;
in the NS sector of the dual $CFT_2$, such short strings states are equivalent to winding one states \refs{\MaldacenaHW\ArgurioTB-\GiveonMI}.}
Plugging \deltadelta\ into \onshellbtz, one finds the dispersion relations
\eqn\elerbtz{E_{L,R}={1\over w}\left[-{j(j+1)\over k}+N_{L,R}-{1\over 2}\right]~.}
The states that satisfy \elerbtz\ can be thought of as describing a string that winds $w$ times around
the spatial circle in the BTZ geometry, and is moving with a certain momentum (proportional to $s$, \jjss)
in the radial direction, in a particular state of transverse oscillation.
Equation \elerbtz\ gives the energy and momentum of such a state.

The spectrum \elerbtz\ is the same as that of ${\cal M}^N/S_N$, where the block $\cal M$
is the $c=6k$ $CFT_2$ associated with one string in the BTZ background, and the winding $w$ labels the twisted sectors.
The string state \letvphys\ corresponds to an operator with dimension $h_w$ in ${\cal M}^w/Z_w$ which,
when acting on the Neveu-Schwartz vacuum of this $CFT_2$, creats a Ramond sector state on the cylinder, with energy
\eqn\elerhh{E_L=h_w-{kw\over 4}~;\quad E_R=\bar h_w-{kw\over 4}~.}

Comments:
\item{*}
Plugging \elerhh\ into \elerbtz, one gets an equation that can be written as
\eqn\hwhonebtz{h_w={h_1\over w}+{k\over 4}\left(w-{1\over w}\right)~,}
which describe the dimensions of the operators in the $Z_w$ twisted sector, $h_w$,
in terms of those in the sector with $w=1$, $h_1$.
A similar equation holds for the right-movers.
\item{*}
\hwhonebtz\ is the expression for the dimension of operators in the $Z_w$ twisted sector
of a symmetric product $CFT_2$, ${\cal M}^N/S_N$, where the block ${\cal M}$ has central charge $c_{\cal M}=6k$.
\item{*}
More comments, which are relevant here, appear already in section 2.

\newsec{$T\bar T$ in perturbative string theory -- heuristic}

%from section 6 in zamo-ws-st ...

Suppose that we have a consistent string theory, say, a perturbative type II superstring,
on a worldsheet $CFT_2$ background that has the following properties:
\item{(1)}
It looks asymptotically like
\eqn\rsrn{R_t\times S^1_R\times R_\phi\times{\cal N}~,}
where the radial direction $\phi$ has an asymptotic linear dilaton $\Phi=-{Q\over 2}\phi=-{1\over\sqrt{2k}}\phi$.
\item{(2)}
In the IR it is $AdS_3\times\cal N$, with the boundary of $AdS_3$ being the cylinder $R_t\times S^1_R$.~\foot{It
implies, in particular, that the theory on the $AdS_3$ cap decouples
from the linear dilaton throat $R_t\times S^1_R\times R_\phi$ when $R/l_s\to\infty$,
where $R/l_s$ is the radius of the circle in string-length units.}

Comments:
\item{(a)}
By `$AdS_3$ whose boundary is the cylinder' we mean either $AdS_3$ in global coordinates or $M=0$ BTZ.
\item{(b)}
For the superstring we need to consider both; the former amounts to the NS sector and the latter to the Ramond sector
of the dual $CFT_2$.
\item{(c)}
${\cal M}_3\times\cal N$, \mthree, is an example of the latter, which will be described in the next section;
it can be obtained from \rsrn\ by adding to it $N$ fundamental strings (F1)
whose worldvolume lies in $R_t\times S^1_R$.~\foot{In the IR,
namely, taking the near-horizon limit of the strings, one gets the geometry of
$M=0\,\, {\rm BTZ}\times{\cal N}$,
and the $F1$'s creating this background turn to the $N$ long strings in the superstring theory on $M=0$ BTZ,
each of which is associated
with a block $\cal M$ of the ${\cal M}^N/S_N$, as was mentioned in sections 2 and 3.}

Heuristically, the physical spectrum of such a string theory is obtained as follows:
\item{*}
Property (1) gives rise, in particular, to~\foot{We shall
use standard perturbative string theory conventions, e.g.,
the string tension is $T=1/2\pi\alpha'$ and the string length scale $l_s$ is related to $\alpha'$ via $\alpha'=l_s^2$;
when $\alpha'$ does not appear, it was set to $\alpha'=2$.}
\eqn\dimvw{-{j(j+1)\over k}+N_{L,R}-{1\over 2}={\alpha'\over 4}\left(E_t^2-p_{L,R}^2\right),}
where $E_t$ is the total energy and
\eqn\plprw{(p_L,p_R)=\left({wR\over\alpha'}+{n\over R},{wR\over\alpha'}-{n\over R}\right)}
is the Narain momentum of a string with winding $w$ and momentum $n$ on the asymptotic circle, $S^1_{x\simeq x+2\pi R}$,
moving with momentum governed by the quantum number $j$ in the radial direction,
in a particular state of transverse left and right-handed levels, $N_{L,R}$;
eq. \dimvw\ is obtained from the mass-shell relation of physical vertex operators whose asymptotic behavior in \rsrn\ is~\foot{We
present only a certain class of observables, for simplicity.}
\eqn\asymptotic{V_{phys}\to e^{-\varphi-\bar\varphi}V_{N_L,N_R}e^{-iE_t t}e^{ip_Lx_L+ip_Rx_R}e^{Qj\phi}~.}
It will also be useful to write the total energy $E_t$ as
\eqn\eeeeee{E_t=E+{wR\over\alpha'}~,}
with $E$ thus being the energy of the state relative to the energy of a BPS string wound $w$ times around the circle
with radius $R$, due to its tension $T$: $2\pi wRT=wR/\alpha'$.
\item{*}
Property (2) gives rise, in particular, to
\eqn\elertt{-{j(j+1)\over k}+N_{L,R}-{1\over 2}=w\left(h_w-{kw\over 4},\bar h_w-{kw\over 4}\right)~,}
for all $w\geq 1$,
where $h_w-kw/4$ ($\bar h_w-kw/4$) is the left-handed (right-handed) energy (in units of $R$)
of the states corresponding to \asymptotic\ in the IR theory, namely, in the $CFT_2$
dual to the superstring theory on the $AdS_3$ cap (with a cylindrical boundary);
eq. \elertt\ is obtained by taking the limit $R/l_s\to\infty$ in \dimvw, which gives rise to
\eqn\elerttt{E_{L,R}(R/l_s\to\infty)\to{1\over w}\left[-{j(j+1)\over k}+N_{L,R}-{1\over 2}\right]~,}
with
\eqn\elrplr{E_{L,R}={1\over 2}(ER\pm n)~,}
and since, on the other hand, property (2) implies that~\foot{For $M=0$ BTZ,
it is presented in \elerbtz,\elerhh, and for global $AdS_3$, it follows from \MaldacenaHW.}
\eqn\elerrinf{E_{L,R}(R/l_s\to\infty)\to\left(h_w-{kw\over 4},\bar h_w-{kw\over 4}\right)~.}
The consequence of \elerttt\ and \elerrinf\ is the equality \elertt.
\item{*}
Together, \dimvw\ and \elertt\ give
${\alpha'\over 4}\left(E_t^2-p_{L,R}^2\right)=w\left(h_w-{kw\over 4},\bar h_w-{kw\over 4}\right)$,
which (with \plprw\ and \eeeeee) we rewrite as
\eqn\honehone{\left(E+{R_w\over\alpha'}\right)^2-\left({R_w\over\alpha'}\right)^2
={2\over\alpha'}\left(h_1+\bar h_1-{k\over 2}\right)+\left({n_w\over R_w}\right)^2~,}
\eqn\hhnw{h_1-\bar h_1=n_w~,}
with $R_w=wR$, $n_w=wn$;
it means that the spectrum of strings with winding $w$ and momentum $n$
is the same as that of a string singly wound around a circle
with radius $R_w$ and momentum $n_w$. This agrees with the spectrum in the $Z_w$ twisted sector of
\eqn\mtnsn{({\cal M}_{-tT\bar T})^N/S_N~; \quad t=\pi\alpha'~,}
with $Z_w$ acting via cyclic permutation on the $w$ copies $\cal M$.

Comments:
\item{(i)}
Note that in the IR limit, $R/l_s\to\infty$,
\honehone\ reduces to the well knwon results in string theory on $AdS_3$
\refs{\MaldacenaHW,\ArgurioTB}, such as \GiveonMI,
\eqn\hhkww{h_w={h_1\over w}+{k\over4}\left(w-{1\over w}\right)~,}
describing long strings winding $w$ times around the boundary circle,~\foot{As it had to, \elertt.}
which is the well known expression for the dimension of operators in the $Z_w$ twisted sector of a symmetric product $CFT_2$,
${\cal M}^N/S_N$, where the block ${\cal M}$ has central charge~\foot{In harmony with the discussion in sections 2 and 3.}
\eqn\cmsixk{c_{\cal M}=6k~.}
\item{(ii)}
More qualitative reasons that support the symmetric product structure are discussed in the previous sections.
\item{(iii)}
For the superstring on ${\cal M}_3\times\cal N$, \mthree,
we proved the relations \honehone,\hhnw\ in a couple of ways \refs{\GiveonMYJ,\ChakrabortyVJA};
some of this will be described in the next two sections.
Here, we point out that
we know \KutasovXU\ that adding $J^-(z)\bar J^-(\bar z)$ to the Lagrangian
of the $AdS_3$ worldsheet theory~\foot{Namely,
to the $SL(2,R)$ WZW model and/or its Euclidean $H_3^+$ version.}
is the same as adding $D(x,\bar x)$ to the Lagrangian of the boundary theory,
\eqn\djj{\int d^2xD(x,\bar x)\simeq\int d^2zJ^-(z)\bar J^-(\bar z)~,}
where the operator $D(x,\bar x)$ has the following properties:
\item{(a)}
$D(x,\bar x)$ transforms under $T(x)$ and $\bar T(\bar x)$ as a quasi-primary operator of dimension $(2,2)$;
its OPE's with $T(x)$ and $\bar T(\bar x)$ is the same as that of $T\bar T$,
\item{(b)}
but $D(x,\bar x)$ is a `single-trace' operator --
it is a massive mode of the dilaton-graviton sector of string theory on $AdS_3$.
\item{*}
These points `dismystify' the result \honehone\--\mtnsn.

\newsec{${\cal M}_3$: $J^-\bar J^-$ deformed $AdS_3$}

Deforming \wssbtz\ by adding to it
%~\foot{The precise coefficient is not important for us here; it can be reabsorbed into the $\lambda$ below.}
\eqn\defbtz{\delta{\cal L}\simeq\lambda J^-\bar J^-~,}
where
$J^-\simeq e^{2\phi}\partial\bar\gamma$ and
$\bar J^-\simeq e^{2\phi}\bar\partial\gamma$
are the null holomorphic and anti-holomorphic currents of the $SL(2,R)$ WZW theory,
one finds \ForsteWP\ a sigma-model background with a metric, dilaton and $B$-field:
\eqn\dsmthree{ds^2=k\left(d\phi^2+fd\gamma d\bar\gamma\right)~,}
\eqn\dilbmthree{e^{2\Phi}=g^2e^{-2\phi}f~,\qquad B_{\gamma\bar\gamma}=kf/2~,}
\eqn\fff{f^{-1}=f_1=\lambda+e^{-2\phi}~,}
which we refer to as ${\cal M}_3$.

From the discussion in section 3, in the representation \lphipm,
the $J^-\bar J^-$ deformation of the sigma model on $AdS_3$ takes the form (at large $\phi$)
\eqn\llbba{{\cal L}=
-\partial\phi_+\bar\partial\phi_- -\partial\phi_-\bar\partial\phi_+ +\lambda\partial\phi_+\bar\partial\phi_+
+\cal L_\phi~;}
the deformation~\foot{The
$\lambda$ in the geometry \fff\ and the one in the representation \llbba\
differ by a factor of $R^2/2\alpha'$, which we ignore here.} thus acts on the two dimensional space labeled by
$(\phi_+,\phi_-)$ by changing the metric $G_{\mu\nu}$ from $\eta_{+-}=-1$, to
\eqn\ggll{G=\pmatrix{
\lambda & -1\cr
-1 & 0\cr
}.}

The geometry of ${\cal M}_3$ depends on the sign of the deformation parameter $\lambda$ in \fff:
\item{*}
For $\lambda>0$, one
finds a smooth asymptotically linear dilaton flat three-dimensional space-time, compactified
on a circle with radius $R$, and capped in the infrared region by a locally $AdS_3$ space;
we will refer to this background as ${\cal M}_3^{(+)}$.
\item{*}
For $\lambda<0$, the background, which we will denote by ${\cal M}_3^{(-)}$, looks as follows. In the
infrared region in the radial coordinate, it approaches $AdS_3$. As one moves towards the UV,
the geometry is deformed, and at some value of the radial coordinate, that depends on $\lambda$,
one encounters a singularity. The region between the IR $AdS_3$ and the singularity looks
like the region between the horizon and the singularity of a black hole. Proceeding past the
singularity, the geometry approaches a linear dilaton spacetime. From the point of view of
an observer living in that spacetime, the singularity in question is naked. Also, the role of
space and time on the boundary are flipped when passing the singularity. Thus, the region
past the singularity has CTC's.
\item{*}
While the backgrounds ${\cal M}_3^{(+)}$ and ${\cal M}_3^{(-)}$
look rather different, their constructions in string
theory are very similar. As described in \GiveonMYJ, the worldsheet theory corresponding to both can
be obtained via null gauging of the worldsheet CFT on $R_t\times S^1_R\times AdS_3$.
For $\lambda>0$ ($\lambda<0$),
the gauging involves an axial (vector) symmetry.
Therefore, it is natural to expect both of them to give rise to good string backgrounds.

\newsec{Superstring theory on ${\cal M}_3$ and $({\cal M}_{-tT\bar T})^N/S_N$}

%from section 5.2 in CGK1806 ...

According to the discussion of sections 2,3 and 4,
fundamental string excitations of the massless BTZ background are described by the symmetric product CFT ${\cal M}^N/S_N$, where $N$ is the number of strings, and ${\cal M}$ the CFT describing one string in this background. One of the interesting aspects of this picture is that, if it is correct,
then the deformation by $J^-\bar J^-$ corresponds to deforming the symmetric product to $({\cal M}_t)^N/S_N$,
where ${\cal M}_t$ is the CFT ${\cal M}$ deformed by the $tT\bar T$ deformation of \refs{\SmirnovLQW,\CavagliaODA}.
Thus, from comment (iii) in section 4,
one expects the string theory analysis of the spectrum to yield in this case the same results as that of the spectrum of
$T\bar T$ deformed CFT in the above papers.

The fact that this is the case was shown in \GiveonMYJ,
using the construction of the ${\cal M}_3$ theory as a null gauging of $R_t\times S^1_R\times AdS_3$.
In this section, following \ChakrabortyVJA,
we will use instead the techniques reviewed in section 3 to derive the same results;
the latter will be useful also for more general theories, e.g. those in sections 7,8 and 9.

\subsec{The spectrum of the worldsheet theory}

The spectrum of a theory with a general constant metric, such as \ggll, is a familiar problem in string theory, in the context of toroidal compactifications, where it gives rise to the Narain moduli space. The slight novelty here is that the deformation involves time, but we can still use techniques developed in the Narain context, and we will do that below.

After the deformation, the scaling dimensions of the vertex operators \vbtz, \vwelr,
which become operators in the sigma model on ${\cal M}_3$, are given by
\eqn\deltaaa{\Delta_{L,R}={1\over 2}P_{L,R}^2-{j(j+1)\over k},}
where~\foot{See e.g. the review \GiveonFU, around (2.4.12) (with $L\leftrightarrow R$).
The antisymmetric background $B$ is zero in the present example;
we keep it for the cases considered in the next sections.}
\eqn\pLpRa{P_{L,R} =\left(n^t+m^t(B\mp G)\right)e^\ast,\qquad P^2\equiv PP^t, \qquad e^\ast(e^\ast)^t ={1\over 2}G^{-1},}
with~\foot{The light-cone momentum is $(n_+,n_-)={1\over\sqrt 2}(n_0+n_1,n_0-n_1)$,
and similarly for the light-cone winding $m$.}
\eqn\nma{\eqalign{n^t& =(n_+,n_-) ={1\over\sqrt 2}\left(2w, \ ER\right),\cr
m^t & =(m_+,m_-) ={1\over\sqrt 2}\left(P R, \ 0\right).}}
Substituting \pLpRa\ and \nma\ in \deltaaa, we get that operators of the form \vbtz,\vwelr\ in ${\cal M}_3$
have left and right scaling dimensions
\eqn\ddbara{\Delta_{L,R}  = -wE_{L,R}-{\lambda R^2\over 8}\left(E^2-P^2\right)-{j(j+1)\over k}~,
\qquad \Delta_R-\Delta_L  = wn~,}
with $E_{L,R}$ given in terms of the energy and momentum $E,P$ and the radius $R$ in \elerp.
This equation generalizes \deltadelta\ to $\lambda\not=0$.

\subsec{The spectrum of the spacetime theory}

We can use the results of subsection 6.1 to calculate the spectrum of the spacetime theory,
as we did in the undeformed case, $\lambda=0$, in section 3.
Using the mass-shell condition \onshellbtz, we find the dispersion relation
\eqn\hbarheea{\eqalign{h_w-{kw\over 4}=& E_L+{\lambda R^2\over 8w}\left(E^2-P^2\right),\cr
\bar{h}_w-{kw\over 4}=& E_R+{\lambda R^2\over 8w}\left(E^2-P^2\right),\cr
h_w-\bar{h}_w=& n,}}
where $h_w$, $\bar h_w$ are properties of the undeformed theory
(e.g., they can be obtained by setting $\lambda=0$ in \hbarheea,
and using the dispersion relations of the undeformed theory, \elerbtz).

It is interesting to compare the spectrum \hbarheea\ to the field theory analysis of $T\bar T$ deformed $CFT_2$ \refs{\SmirnovLQW,\CavagliaODA}.
It is easy to see that the two agree, if we take the boundary $CFT_2$ to be the symmetric product ${\cal M}^N/S_N$,
interpret the deformation \llbba\ to be the $T\bar T$ deformation in ${\cal M}$,
and take the coupling $\lambda$ in the string theory problem to be related to the $tT\bar T$ coupling, $t$, via
\eqn\tlrr{t= {\pi\over2}\lambda R^2.}
For $w=1$, the spectrum \hbarheea\ is just that of $T\bar T$ deformed $CFT_2$ for the deformed block $\cal M$,
${\cal M}_t$.
For $w>1$, it is that of the $Z_w$ twisted sector of the symmetric product $({\cal M}_t)^N/S_N$~\GiveonMYJ.

Comments:
\item{*}
The energy of states in the $tT\bar T$ deformed $CFT_2$ of the
block ${\cal M}_t$ takes the form (for $P=0$, namely, $\bar h-h=n=0$, for simplicity):
\eqn\etrcm{E(t,R)={\pi R\over t}\left[-1+\sqrt{1+{4t\over\pi R^2}\left(h-{c_{\cal M}\over 24}\right)}\right]~.}
Some properties of the spectrum are the following:
\item{*}
In the IR, namely, for $h-{c_{\cal M}\over 24}\ll{R^2\over t}$,
the energy \etrcm\ is parametrically close to that of the CFT $\cal M$,
$ER\simeq 2\left(h-{c_{\cal M}\over 24}\right)$, as it should.
\item{*}
On the other hand, in the UV, when $t>0$, the large energy behavior of \etrcm\ is
$E\simeq\sqrt{{4\pi\over t}\left(h-{c_{\cal M}\over 24}\right)}$,
hence, the entropy,~\foot{We use the fact that the degeneracy of states does not change when we turn on $t$,
and thus $S(E(h))=S(h)$, where the latter is the Cardy entropy of the CFT $\cal M$.}
\eqn\entropy{S((E(h))\simeq 4\pi\sqrt{{c_{\cal M}\over 6}\left(h-{c_{\cal M}\over 24}\right)}\simeq\sqrt{2\pi c_{\cal M}t\over 3}E
=\beta_H E~,}
is a Hagedorn one, with $\beta_H$ being
the circumference of the circle, $2\pi R$, at the point where the
ground state energy, $E(h=0)$, becomes complex; see \GiveonNIE\ for a detailed discussion.
\item{*}
The high energy behavior of the entropy of the deformed
symmetric product CFT, $({\cal M}_t)^N/S_N$, was shown~\foot{Using the fact that the maximal entropy
configuration in $({\cal M}_t)^N/S_N$ is one in which the total energy $E$ is divided equally in each of the $N$ blocks,
namely, $S_{total}(E)=NS_{\cal M}(E/N)$; see \GiveonNIE\ for more details.}
to agree with the Bekenstein-Hawking entropy of black
holes in the deformed geometry induced by the single-trace deformation, ${\cal M}_3^{(+)}$.
\item{*}
From the holographic duality above, the Hagedorn behavior is clear,
since the UV completion provided by string theory on ${\cal M}_3^{(+)}$
is a $2d$ Little String Theory.
\item{*}
When $t<0$, there is a maximal value, $E_{max}={\pi R\over|t|}$, above which the energies develop an imaginary piece.
\item{*}
Many aspects of the discussion of the partition sum in \AharonyBAD\ have a natural interpretation in the above
string theory construction:
\item{($+$)}
For example, we found that for $t>0$, the spectrum of the $tT\bar T$ deformed
theory does not receive non-perturbative corrections. This is natural in the string theory
construction since ${\cal M}_3^{(+)}$ is a smooth space. The explicit calculation above shows that the states in
string theory on ${\cal M}_3^{(+)}$ described by the symmetric product do indeed have a smooth limit
as $t\to 0^+$.
\item{($-$)}
On the other hand, for $t<0$, we found that the partition sum of the theory has a
non-perturbative ambiguity, which corresponds to states with
energies that diverge as $t\to 0^-$.
It would be interesting to understand these and other
features of the field theory discussion from the string theory perspective.
It is tempting to speculate that states whose energies have a good perturbative limit
correspond in the bulk to wavefunctions that in some sense live in the region between the horizon and the singularity,
while those whose energies diverge in the limit $t\to 0$ live in the region beyond the singularity.
Analyzing this could shed light on whether the singularity of the space-time ${\cal M}_3^{(-)}$
is resolved in string theory, and how. We hope to return to this subject in future work.~\foot{The above comments suggest, however,
that the other branch of the solutions to the quadratic equation for $E$, namely, \etrcm\ with a minus in front of the square root,
is unambiguous in string theory.}

\newsec{$WAdS_3$: $K\bar J^-$ deformed $AdS_3\times S^1$}

%from section 3 in CGK1806 ...

The worldsheet sigma-model Lagrangian on $M=0$ BTZ$\times S^1$ is~\foot{I
am very telegraphic here; more details can be found in \ChakrabortyVJA\ and references therein.}
\eqn\wssbtzs{{\cal L}=2k\left(\partial\phi\bar\partial\phi+e^{2\phi}\partial\bar\gamma\bar\partial\gamma
+\partial y\bar\partial y\right)~,}
with a compact $\gamma_1$, \gggggr. Deforming \wssbtzs\ by adding to it
%~\foot{The precise coefficient is not important for us here; it can be reabsorbed into the $\epsilon$ below.}
\eqn\ldefkj{\delta{\cal L}\simeq\epsilon K\bar J^-~,\qquad K\simeq i\partial y~,}
and KK  reducing the resulting sigma-model background to $3d$, one finds the geometry
\eqn\wadsgeom{ds^2=k\left(d\phi^2+e^{2\phi}d\gamma d\bar\gamma -\epsilon^2 e^{4\phi}d\gamma^2\right)~,}
with a gauge field, $A_\gamma=2\sqrt{k}\epsilon e^{2\phi}$, and a $B$-field, $B_{\gamma\bar\gamma}=ke^{2\phi}/2$.

In the representation \lphipm, the $K\bar J^-$ deformation of the sigma model on $AdS_3$ takes the form (at large $\phi$)
\eqn\Wlagf{{\cal{L}}=-\partial\phi_+\bar{\partial}\phi_--\partial\phi_-\bar{\partial}\phi_+ +\partial y\bar{\partial}y+2\epsilon\partial y\bar{\partial}\phi_+ +{\cal L}_\phi~;}
this background~\foot{The $\epsilon$ in the geometry \wadsgeom\ and the one in the representation \Wlagf\
differ by a factor $R/\sqrt 2l_s$, which we ignore here.}
involves a non-trivial metric and $B$-field background in the three dimensional spacetime labeled by $(\phi_+,\phi_-,y)$,
\eqn\GB{G=\pmatrix{
0 & -1& \epsilon \cr
-1 & 0 & 0\cr
\epsilon & 0 & 1
}, \ \ \ \ B=\pmatrix{
0 & 0 & -\epsilon \cr
0 & 0 & 0 \cr
\epsilon & 0 & 0
}.}

Comments:
\item{*}
The background \wadsgeom\ is the simplest example of warped $AdS_3$ ($WAdS_3$);
it appears e.g. as the simplest example in the context of AdS/cold atoms correspondence \refs{\SonYE,\BalasubramanianDM},
and may have applications as a toy laboratory for the Kerr/CFT correspondence \AzeyanagiZD.
\item{*}
The geometry \wadsgeom\ has no curvature singularity and, in particular, the scalar curvature is an $\epsilon$-independent constant,
${\cal R}\simeq -1/k$; there are, however, CTC's at large values of the radial coordinate, when $e^{2\phi}>1/\epsilon^2$.
\item{*}
We know \KutasovXU\ that adding $K(z)\bar J^-(\bar z)$ to the Lagrangian of the worldsheet theory
is the same as adding $A(x,\bar x)$ to the Lagrangian of the boundary theory,
\eqn\djj{\int d^2xA(x,\bar x)\simeq\int d^2zK(z)\bar J^-(\bar z)~,}
where the operator $A(x,\bar x)$ has the following properties:
\item{(a)}
$A(x,\bar x)$ has dimension $(1,2)$.
\item{(b)}
It transforms under $J(x)$ and $\bar T(\bar x)$ of the boundary theory dual to string theory on $AdS_3\times S^1$
like $J\bar T$,
\item{(c)}
but $A(x,\bar x)$ is a `single-trace' operator -- it is a marginal operator in the worldsheet theory.

In the following section, we shall compute the spectrum of the superstring on the $WAdS_3$ background \wadsgeom,
and thus `predict' the spectrum of $J\bar T$ deformed $CFT_2$.

\newsec{Superstring theory on $WAdS_3$ and $({\cal M}_{\mu J\bar T})^N/S_N$}

%from section 5.3 in CGK1806 ...

The analogs of \vbtz, \vwelr\ for this case are
vertex operators $V$ in the deformed massless BTZ$\times S^1$ background,
which are obtained by multiplying \vbtz\ by an extra factor of $e^{i(q_L y+q_R\bar y)}$,
where $(q_L,q_R)$ is a Narain momentum on the $S^1_y$.~\foot{See section 5.3 in \ChakrabortyVJA\ for details.}
Using the standard techniques, described in section 6.1, one finds that the left and right-handed scaling dimensions of $V$ are
\eqn\ddbarab{\Delta_{L,R} = -wE_{L,R}+\epsilon q_LE_R+{1\over 2}\epsilon^2 E_R^2+{1\over 2}q_{L,R}^2-{j(j+1)\over k}~,
\quad \Delta_R-\Delta_L  ={1\over 2}\left(q_R^2-q_L^2\right)+wn~,}
and repeating similar steps to those in section 3 and subsection 6.2, one finds that the spectrum of the spacetime theory is
given by the dispersion relation
\eqn\hbarheeab{\eqalign{h_w-{kw\over 4}=& E_L-{\epsilon\over w}q_LE_R-{\epsilon^2\over 2w}E_R^2,\cr
\bar{h}_w-{kw\over 4}=& E_R-{\epsilon\over w}q_LE_R-{\epsilon^2\over 2w}E_R^2,\cr
h_w-\bar{h}_w=& n.}}

Comments:
\item{*}
To understand the physical content of \hbarheeab, it is convenient to rewrite it as follows.
Consider first the case $w=1$, corresponding to a string with winding one.
From the discussion in earlier sections, in this sector we expect to see the spectrum of a $J\bar T$ deformed $CFT_2$
with central charge $c=6k$ (before the deformation).
In this sector, one can write \hbarheeab\ as follows:
\eqn\hcqq{h_1-{c\over 24}-{1\over 2}q_L^2=E_L(\epsilon)-{1\over 2}(q_L(\epsilon))^2,\quad
E_L-E_R=n,}
where $c=6k$, and
\eqn\qller{q_L(\epsilon)=q_L+\epsilon E_R(\epsilon).}
We see that the spectrum is completely determined by the following requirements:
\item{(1)} The quantity $E_L(\epsilon)-{1\over 2}(q_L(\epsilon))^2$ is independent of $\epsilon$.
\item{(2)} The charge $q_L$ flows according to \qller.
\item{(3)} For all $\epsilon$, $E_L$ and $E_R$ differ by the second eq. in \hcqq.
\item{(4)} At $\epsilon=0$, one has $q_L(0)=q_L$, $E_L(0)=h_1-{c\over 24},\, E_R(0)=\bar h_1-{c\over 24}$.
\item{*}
In a $\mu J(x)\bar T(\bar x)$ deformed $CFT_2$, one obtains \refs{\ChakrabortyVJA,\AharonyICS} the same spectrum, with
\eqn\mulamr{\mu=2\epsilon R.}
\item{*}
For $w>1$, the spectrum \hbarheeab\ is compatible with that of a $Z_w$ twisted sector of the symmetric product $({\cal M}_\mu)^N/S_N$,
where the block ${\cal M}_\mu$ is deformed as described above.
\item{*}
The energy of states in the $w=1$ sector takes the form (for $n=q_L=0$, for simplicity)
\eqn\ereeh{E(\mu,R)={8R\over\mu^2}\left[1-\sqrt{1-{\mu^2\over 2R^2}\left(h-{c_{\cal M}\over 24}\right)}\right]~,}
where $h$ is the dimension of the operator which creates the state in the undeformed $CFT_2$ block $\cal M$.
\item{*}
There is a maximal value of this energy, $E_{max}={8R\over\mu^2}$,
above which the energies develop an imaginary piece.~\foot{Also for generic momentum and charge, $n,q_L$,
beyond a certain maximal undeformed (right-handed) energy, $\bar h-c_{\cal M}/24$, that depends
on the charge, the deformed energy and charge become complex
(this can be seen explicitly e.g. from equations presented in section 9.2).}
\item{*}
The spectrum \ereeh\ is identical to that of the $tT\bar T$ deformed case \etrcm\ with a negative deformation parameter,
$t=-{\pi\mu^2\over 8}$.~\foot{The
spectra are different though for non-zero momentum $P=n/R$ and/or for non-zero $U(1)$ charge $q_L$,
if only since the $J\bar T$ deformed theory is {\it not} Lorentz invariant.}
\item{*}
It is thus interesting to compare the properties of the boundary and bulk theories in the $T\bar T$
and $J\bar T$ cases. As discussed in \AharonyICS,
on the field theory side, many properties of $J\bar T$ deformed
CFTs are indeed analogous to those of a $tT\bar T$ deformed CFT with negative $t$.
In particular, not only does the energy spectrum becomes complex in the UV,
but also the partition sum has non-perturbative ambiguities in both cases.
On the bulk side of the ST-$tT\bar T$ case,
as discussed in section 5, the background has a
curvature singularity at a finite value of the radial coordinate, and closed timelike curves
beyond it. In the ST-$J\bar T$ case, as discussed in section 7,
there is no curvature singularity, but there are closed timelike
curves at large values of the radial coordinate.
Thus, it is natural to conjecture that
the complex energies and non-perturbative ambiguities mentioned above are related to the
closed timelike curves and not to the curvature singularity.~\foot{New evidence
in support to this claim is presented in the next section.}
\item{*}
It would be interesting to understand this relation better, and in particular understand
whether the theory is well defined on the cylinder after all,
despite the issues with complex energies, non-perturbative ambiguities and closed timelike curves.
One possible way to go about this is to further explore
the string theory formulation of the theory, as a current-current deformation of string theory
on $AdS_3\times S^1$; we leave this too for future work.

\newsec{Combining the two: spectrum versus gometry and $({\cal M}_{-tT\bar T+\mu J\bar T})^N/S_N$}

%some preliminary results (from ttjt file) ...

In this section, we deform the massless BTZ$\times S^1$ Lagrangian, \wssbtzs,
by adding to it
\eqn\delcom{\delta{\cal L}=\tilde\lambda J^-\bar J^-+\tilde\epsilon K\bar J^-~,}
namely, \defbtz\ plus \ldefkj, inspect the $3d$ geometry obtained (after KK reduction)
and the spectrum of the superstring on this background.
The results, presented below,
are the outcomes of straightforward, simple calculations,
following what was done in sections 3,5,6,7,8 and, in particular,
they predict the spectrum of $-tT\bar T+\mu J\bar T$ deformed $CFT_2$.

\subsec{The $3d$ geometry holographic to single-trace $-tT\bar T+\mu J\bar T$}
The metric, dilaton, gauge field and $B$-field of the sigma-model background \wssbtzs\
deformed by \delcom,
after reduction to three dimensions, are~\foot{The precise factors between $\tilde\lambda,\tilde\epsilon$ in \delcom\ and $\lambda,\epsilon$, respectively, are not important here.}
\eqn\geometrycom{ds^2=k\left(d\phi^2+fd\gamma d\bar\gamma-\epsilon^2f^2 d\gamma^2\right)~,}
\eqn\dilatoncom{e^{2\Phi}=g^2e^{-2\phi}f~,\quad A_\gamma=2\sqrt k\epsilon f~,\quad B_{\gamma\bar\gamma}=kf/2~;}
\eqn\fffcom{f^{-1}=f_1=\lambda+e^{-2\phi}~.}
Comment:
\item{*}
The $\lambda,\epsilon$ (in the geometry) are proportional to the $t,\mu$ (in the field theory), respectively:~\foot{These
precise relations can be read from \refs{\GiveonMYJ,\ChakrabortyVJA}.}
\eqn\tlme{t=\pi\alpha'\lambda~,\quad \mu=2\sqrt 2\ell_s\epsilon~;\quad \alpha'=\ell_s^2~.}

In the representation \lphipm,
the metric and $B$-field in the three dimensional space $(\phi_+,\phi_-,y)$ are given by the combination
of \ggll\ and \GB, namely,
\eqn\GBcom{G=\pmatrix{
\hat\lambda & -1& \hat\epsilon \cr
-1 & 0 & 0\cr
\hat\epsilon & 0 & 1
}, \ \ \ \ B=\pmatrix{
0 & 0 & -\hat\epsilon \cr
0 & 0 & 0 \cr
\hat\epsilon & 0 & 0
}.}
The $\lambda,\epsilon$ (in \geometrycom\--\fffcom) are related to the hatted ones (in \GBcom) by~\foot{The
precise relation can be read from \ChakrabortyVJA.}
\eqn\lehatle{\{\lambda,\epsilon^2\}={R^2\over 2\alpha'}\{\hat\lambda,\hat\epsilon^2\}~.}

\subsec{The spectrum}
The spectrum of the superstring on the background in the previous subsection is obtained straightforwardly
following the discussion in the previous sections. One finds  the dispersion relation
\eqn\hhhcom{h_w-{cw\over 24}=E_L-{\hat\epsilon\over w}q_LE_R+{1\over 2w}(\hat\lambda E_L-\hat\epsilon^2E_R)E_R~,}
\eqn\barhcom{h_w-\bar h_w=E_L-E_R=n~,}
hence, for $w=1$ (for simplicity)
\eqn\eee{ER=E_L+E_R=n+{1\over 2A}\left(-B-\sqrt{B^2-4AC}\right)~,}
with~\foot{The branch of the two solutions to the quadratic equation for $E$, that follows from \hhhcom,\barhcom,
which is presented in \eee, is the one connected to the $CFT_2$ spectrum; on the other hand,
the other branch either decouple or is unphysical when the deformation is turned
off: $E(\lambda\to 0^\pm,\epsilon\to 0)\to \pm\infty$ when $A\to 0^\pm$, where $A={\alpha'\over 2R^2}(\epsilon^2-\lambda)$.}
\eqn\abc{A={1\over 4}(\hat\epsilon^2-\hat\lambda)~,\quad B=-1+\hat\epsilon q_L-{1\over 2}\hat\lambda n~,\quad C=2\bar h_1-{c\over 12}~;}
\eqn\csixk{n=h_1-\bar h_1~,\qquad c=6k~,}
for instance, when $n=q_L=0$ (for simplicity), the energy of states in the $w=1$ sector takes the form
\eqn\eeel{E(\lambda,\epsilon;R)={R\over\alpha'\left(\epsilon^2-\lambda\right)}
\left[1-\sqrt{1-{4\alpha'\over R^2}\left(\epsilon^2-\lambda\right)\left(h_1-{c\over 24}\right)}\right] ~.}

Comments:
\item{*} The dispersion relation \hhhcom,\barhcom\ is the minimal combination of \hbarheea\ and \hbarheeab,
which reduces to them when either $\epsilon=0$ or $\lambda=0$.
\item{*}
The structure of the spectrum is, again, as in a symmetric product ${\cal M}^N/S_N$,
with the spectrum in the block $\cal M$ given by \eee\--\csixk.
\item{*}
{\it We are thus led to conjecture that \eee\--\csixk\
is the spectrum of a $-tT\bar T+\mu J\bar T$ deformed $CFT_2$, ${\cal M}$};~\foot{For
$n=0$, this was already verified (with reasonable assumptions) \cdgjk.}
for instance, when $n=q_L=0$ (for simplicity), the energy is conjectured to take the form
\eqn\etmu{E(t,\mu;R)={8\pi R\over\pi\mu^2-8t}
\left[1-\sqrt{1-{1\over 2\pi R^2}\left(\pi\mu^2-8t\right)\left(h-{c_{\cal M}\over 24}\right)}\right]~.}
\item{*}
Note that at large $\phi$, the part of the metric \geometrycom\ in the $\gamma,\bar\gamma$ directions behaves like
$(\lambda d\bar\gamma-\epsilon^2d\gamma)d\gamma$,~\foot{And recalling that $\bar\gamma,\gamma$ are conjugate to $E_{L,R}$, respectively,
this is in harmony with the factor $(\hat\lambda E_L-\hat\epsilon^2 E_R)E_R$ in the spectrum \hhhcom.}
hence, $\gamma_1$ is spacelike at large $\phi$ if $\lambda-\epsilon^2>0$,
but if $\lambda-\epsilon^2<0$, then
$\gamma_1$ is {\it timelike} at large $\phi$.~\foot{Note that
the determinant of the metric \geometrycom\ is $\epsilon$-independent and, in particular,
there is thus one time for any $\epsilon$ (and $\phi$),
but which of the directions is timelike does depend on $\epsilon$ (and $\phi$).}
\item{*}
One can thus see that the background \geometrycom\ has closed timelike curves
precisely when the spectrum \eee,\abc\ has the property that
beyond a certain maximal undeformed (right-handed) energy, $\bar h_1-c/24$, that depends
on the momentum and charge, $n,q_L$, the deformed energy become complex.
\item{*}
In particular, when e.g. $n=q_L=0$,
there is a maximal value of the deformed energy, $E_{max}={R\over\alpha'(\epsilon^2-\lambda)}$,
above which the energies in \eeel\ develop an imaginary piece.
\item{*}
All in all, {\it we are thus led to conjecture that complex energies in the UV
are associated with closed timelike curves in the holographic string-theory geometry}.
%\item{3'.} To understand the interpretation of complex energies in the parameters space
%           (whether non-unitarity or non-locality associated w/ Hagedorn physics)
%           we should 1st write down the energy $E(\epsilon,\lambda,q)$ from eqs. \hhh,\barh\
%           (the analog of (3.6) in our last paper) to see its precise structure,
%           and then think about its meaning.
%\item{4.} I have not checked the results here carefully yet..
%\item{5.} Recall that, as in eqs. (5.39),(5.40) of the `$J\bar T$ and string theory' paper,
%          eq. \hhhcom\ above can be written as
%\eqn\hhqq{h-{c\over 24}-{1\over 2}q^2=E_L+{1\over 2}\hat\lambda E_LE_R-{1\over 2}(q(\hat\epsilon))^2}
%with
%\eqn\qqq{q(\hat\epsilon)=q+\hat\epsilon E_R}
%where $q(\hat\epsilon)$ is the value of the charge w.r.t. the affine $U(1)$ in the $\lambda=0$ cases,
%and $q$ is its value in the original $CFT_2$.

\newsec{Open problems}

\item{(i)}
Is the dual theory indeed ${\cal M}^N/S_N$, or what precisely?~\foot{We
find non-trivial match and confirmed predictions if we {\it assume} ${\cal M}^N/S_N$;
but what is the {\it precise} duality?}

\item{(ii)}
Does the dual theory tell us if/how {\it string theory resolves singularities/CTC's}?
\item{(iii)}
Does string theory tell us what is the {\it definition} of $tT\bar T$ deformed $CFT_2$
(for both signs of $t$) and the other irrelevant deformations above?
\item{(iv)}
Applications to AdS/cold atoms and Kerr/CFT correspondence?
\item{(v)}
Modular invariance in general cases?
\item{(vi)} Is the conjecture in section 9 correct?

\bigskip\bigskip
\noindent{\bf Acknowledgements:}
The main results and comments in this note are based on
\refs{\GiveonNIE,\GiveonMYJ,\ChakrabortyVJA,\AharonyBAD,\AharonyICS}
and the work in progress \cdgjk,
in collaboration with O.~Aharony, S.~Chakraborty, S.~Datta, N.~Itzhaki, Y.~Jiang and D.~Kutasov.
This work is supported in part by a center of excellence supported by the Israel Science Foundation
(grant number 2289/18).

\listrefs
\end